\documentclass[sigconf]{acmart}
\AtBeginDocument{%
  }

\setcopyright{acmlicensed}
\copyrightyear{2026}
\acmYear{2026}
\acmDOI{XXXXXXX.XXXXXXX}
\begin{document}

\title{From Clicking to Moving: Embodied Micro-Movements as a New Modality for Data Literacy Learning}
\author{Annabella Sakunkoo}
\affiliation{%
  \institution{Stanford University OHS}
  \country{USA}
}
\email{apianist@ohs.stanford.edu}


\author{Jonathan Sakunkoo}
\affiliation{%
 \institution{University of Oxford, Mathematical Institute and Computer Science Department}
  \city{Oxford}
  \country{UK}
}
\email{jonathan.sakunkoo@cs.ox.ac.uk}



\begin{abstract}
Widespread digital learning has expanded access to education but has resulted in highly sedentary, click-based interaction, contributing to digital fatigue, reduced cognitive flexibility, and health risks associated with prolonged passive screen time. Meanwhile, data literacy has become an essential competency in a data-driven society, yet it is typically taught through passive, disembodied interfaces that offer little physical engagement. We present Kinetiq (Kinetic+IQ), a novel system that integrates fun, full-body micro-movements directly into data and numeracy problem solving. Instead of selecting answers with a mouse, learners interact through natural gestures such as reaching, dodging, heading, elbowing, or knee-raising, thus turning abstract data problem-solving into embodied experiences that integrate thinking with movement. In a preliminary within-subjects study comparing Kinetiq with conventional platforms, participants reported significantly higher affective valence, enjoyment, engagement, and motivation, while maintaining comparable learning gains. We contribute: (1) a task-integrated movement paradigm for data learning, (2) a cross-platform web and mobile app system enabling full-body learning in constrained everyday spaces, and (3) preliminary empirical evidence that embodied micro-movements can enrich the affective experience of data literacy learning.

\end{abstract}

\begin{teaserfigure}
\centering
  \includegraphics[width=0.6\textwidth]{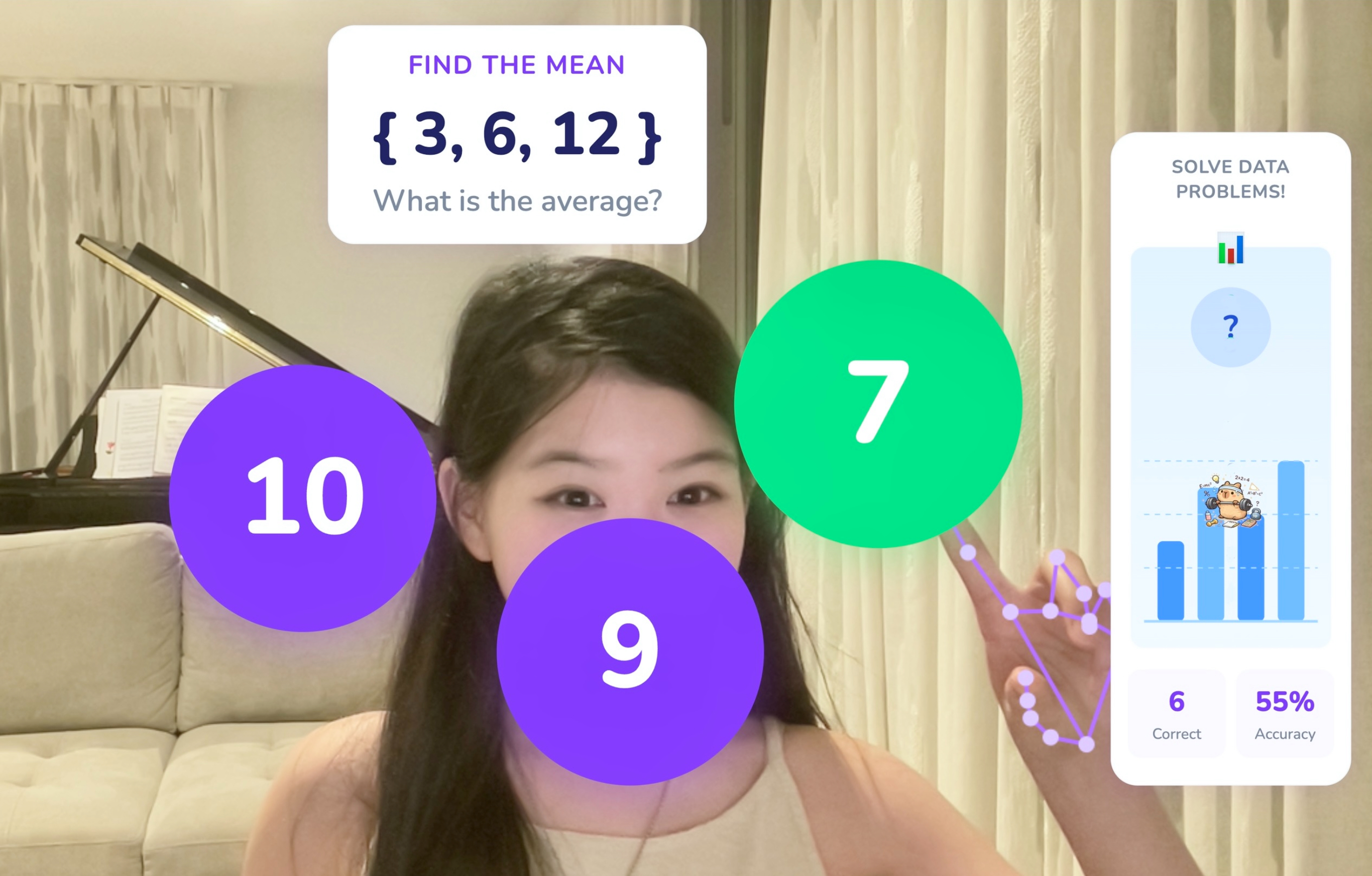}
 {}
 \caption{Kinetiq for Data Literacy Learning with Active Movement}
  \Description{Enjoying the baseball game from the third-base
  seats. Ichiro Suzuki preparing to bat.}
  \label{Data1.jpeg}
\end{teaserfigure}
\maketitle
\section{Introduction}

Data literacy, the ability to read, work with, analyze, and reason with data, has become a foundational competency for navigating the modern world. Yet despite its growing importance, prior work shows that statistics and data literacy are often perceived by learners as unengaging, tedious, or intimidating, perceptions that result in disengagement and avoidance even when students recognize their practical value \cite{Meng}.
At the same time, digital platforms such as Khan Academy, IXL, and Quizlet have dramatically expanded access to data analysis and statistics education. While these tools have successfully scaled content delivery, they rely almost exclusively on sedentary, screen-based interaction, primarily clicking, tapping, and watching. Consequently, students now spend much of their day in prolonged physical stillness, with over seven hours in sedentary activities on average \cite{harvardMuchSit}. Education researchers have found that many students experience school and online learning as boring, tiring, and demotivating \cite{harvard2017}. Prior research also links extended sedentary behavior to cognitive and emotional challenges, including reduced executive function and diminished wellbeing \cite{who2020}.
Together, these trends reveal a persistent tension in digital data education: learners are typically asked to improve their data literacy through environments that minimize physical engagement. This disconnect suggests an opportunity to rethink how data learning environments are designed, not only in terms of content, but in how learners learn and practice.
We introduce Kinetiq, a web-based “move and learn” system that reframes data learning through full-body interaction. Instead of selecting answers with a mouse or touchscreen, learners physically reach, strike, or dodge using different parts of their body. Kinetiq draws on metaphors from the natural world to support intuitive, exploratory engagement while integrating cognitive training with physical movement. Using computer vision and requiring no specialized hardware or installation, the system functions even in small spaces.
In a preliminary user study, we find that Kinetiq substantially improves enjoyment, focus, and motivation compared to conventional digital learning tools. These results suggest that embodied interaction can offer a complementary pathway to supporting data literacy and mathematical understanding, one that engages both mind and body in everyday learning environments.

\section{Related Literature}
 Data and statistics literacy has become critical competencies in a data-driven society. Yet it remains one of the most challenging subjects for many learners \cite{Onwuegbuzie2003AssessmentIS}. Meng (2009) documents common student reactions such as “statistics was the most boring course I took” and “that was really a hard course for me,” suggesting that it is frequently undervalued because conventional pedagogical approaches fail to make learning engaging \cite{Meng}.
In parallel, a growing body of work has examined how digital tools and interactive systems can support data and statistics literacy \cite{Burckhardt01012021}. While these platforms have expanded access to learning resources, many students still describe traditional digital learning environments as "depressing," "useless," and even a "torture" \cite{Trustpilot}. Moreover, most of these systems rely on sedentary, screen-centered interaction. Average screen time has risen markedly in recent years \citep{smith2024designing, duran2023breaking}, and research links prolonged sedentary behavior to declines in mood, attention, and perceived wellbeing \cite{mcgonigal2019}. Together, these trends motivate the design for data literacy learning environments that move beyond passive, click-based interaction.
A separate line of research demonstrates that physical action can meaningfully support cognition. Empirical work has demonstrated that thinking is deeply intertwined with bodily experience \citep{wilson2002,shapiro2025} that can play a productive role in learning rather than simply serving as a break from it \cite{novack2015}. In educational contexts, studies show that integrating movement directly with instructional content, rather than treating it as a separate activity before or after learning, can improve executive function and retention \citep{schmidt2016executive, vazou2020cognition}. These benefits appear strongest when movement is task-relevant and temporally aligned with the learning goal, such that physical action and cognitive activity occur together \cite{vazou2020cognition}. 
However, prior systems that incorporate movement into learning often treat physical activity as ancillary to cognition or require specialized hardware, large spaces, or wearables that limit everyday use. As a result, they are rarely integrated into mainstream digital learning environments for data or statistics education.
Recent work has begun to explore tangible approaches to learning statistical concepts \cite{Fan2024}, but the intersection of body-based interaction and data literacy learning remains largely underexplored. Kinetiq builds on this literature by embedding data reasoning directly within full-body interaction, offering a low-barrier, innovative approach that unites movement and sensemaking in the same learning moment. In doing so, it proposes a new modality for developing data competencies that is both more engaging and more intuitively grounded than conventional screen-based systems.

\begin{figure}[h]
  \centering
\includegraphics[width=0.55\linewidth]
{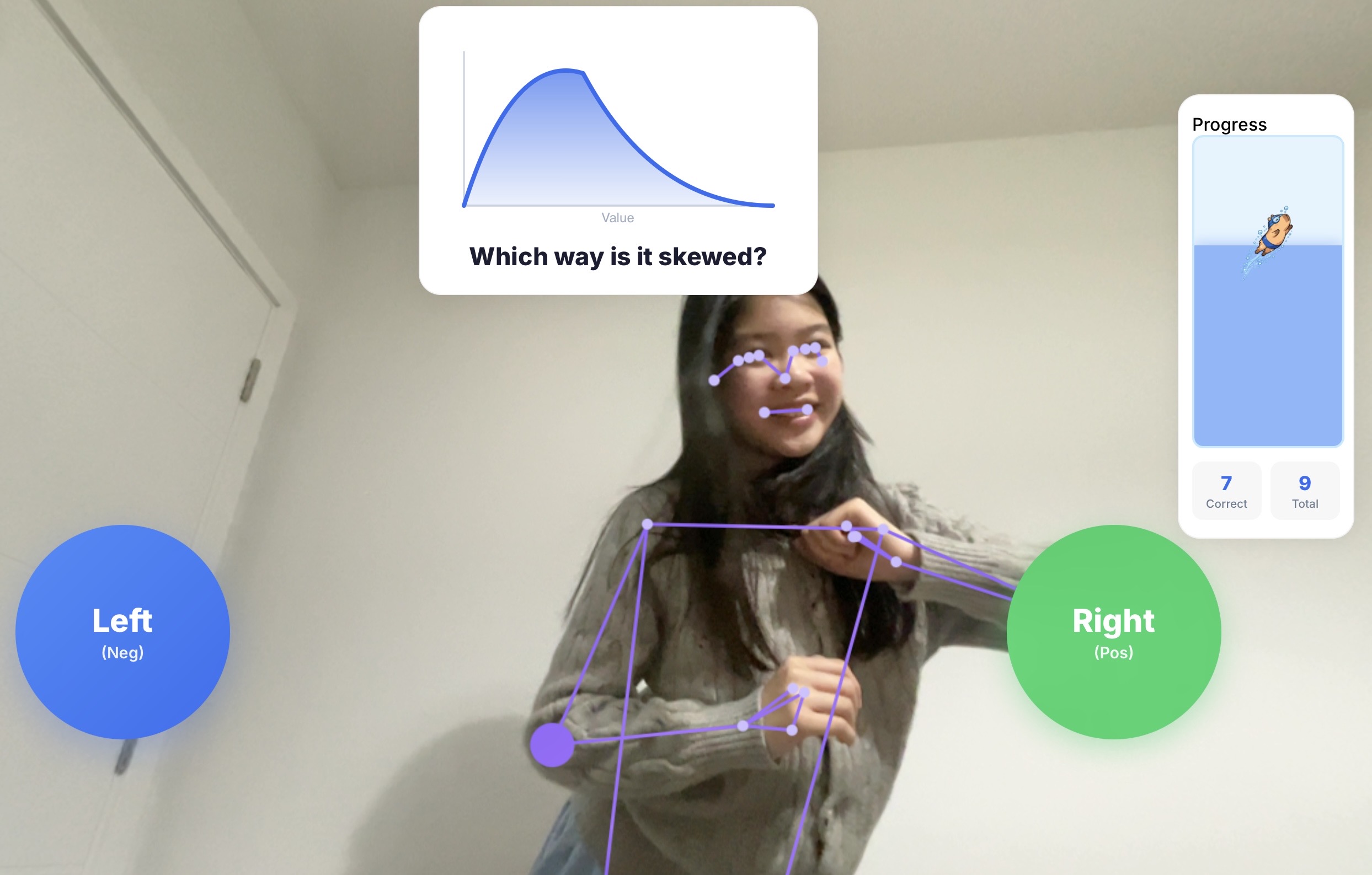}
  \caption{Kinetiq transforms sedentary digital learning into a physically interactive experience using computer vision. It allows users to do data literacy practices with full-body gestures and hence makes learning more engaging, active, and joyful while improving motivation. The user here extends an arm and moves the body, meaningfully using the "right" elbow to hit the Right-Skewed answer.}
\end{figure}

\section{System Description}

\subsection{Design}
Kinetiq addresses these gaps through a webcam-only, zero-installation, small-space system that embeds micro-movements directly within cognitive training. The system features three characteristics to promote playful learning: a constructive view on failure, promoting engaging immersion, and cultivating motivation for learning \cite{Li2024}. 
Our interaction design is also informed by the Mechanics–Dynamics–Aesthetics (MDA) framework \cite{hunicke2004mda}, a model for designing engaging and meaningful player experiences. 
\subsubsection{Aesthetics}
The aesthetic goal of Kinetiq is to evoke feelings of joy and vitality through playful movement. It also evokes feelings of accomplishment, via progressive narratives (e.g. climbing, swimming), mindfulness, by encouraging presence through bodily awareness and task alignment, and engagement and immersion, through multisensory feedback. Kinetiq’s visuals were designed to be both expressive and readily interpretable, drawing on established visualization principles \cite{Mackinlay}. Our design presents simple visual features as prior work on the study of fun has found that visually simpler puzzle designs are more liked by users \cite{Chu2025}. 

\subsubsection{Dynamics}
The dynamics emerges from player-system interaction. Kinetiq balances challenge and flow as it allows players to self-regulate intensity through movement, speed, and rhythm. By integrating custom music selection and offering both solo and multiplayer modes, the system supports personalization and social engagement of cooperative dynamics. Time constraint is also implemented for challenges.
\begin{figure}[h]
  \centering
  \includegraphics[width=0.5\linewidth]
  {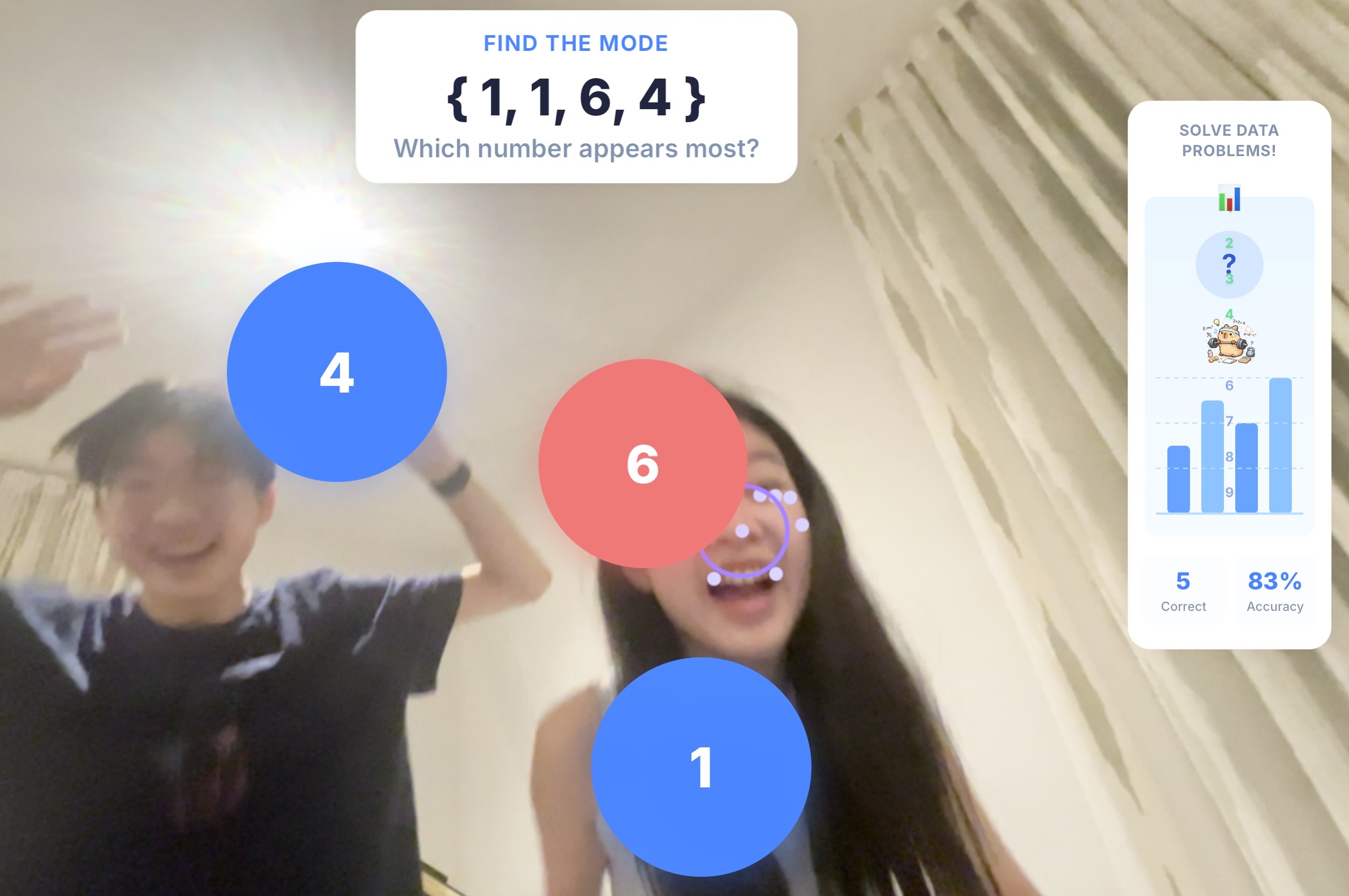}
  \caption{Collaborative, fun learning as users move, stretch, and dodge while trying to bump correct answers with their `heads'.}
  \Description{A woman and a girl in white dresses sit in an open car.}
\end{figure}
\subsubsection{Mechanics}
Kinetiq features real-time pose estimation to recognize gestures such as hitting, elbowing, leaning, heading, and knee raises. The system runs entirely on-device in the user's browser and processes frames at 25–30 FPS on standard consumer hardware with a typical gesture recognition latency under 100 ms. No webcam data is transmitted to or stored on external servers. It sets aside body landmarks that are occluded or outside the camera frame and applies smoothing to reduce jitter between frames. As a cross-platform embodied learning system for constrained everyday spaces via webcam or mobile camera, the system is designed with tolerance for variation to ensure accessibility, even for users with limited mobility or space.

To enhance player engagement, visual and auditory feedback is synchronized with on-screen events. These multimodal feedback strategies follow principles from game design, which emphasize contingent, real-time reinforcement to support motivation, learning retention, and flow \cite{habgood2011motivation}. The overall structure supports both mastery-oriented learning and sustained player interest through a continuous cycle of challenge, action, and reinforcement.

Kinetiq’s visual design also aims to be both expressive and effective as informed by visualization design principles \cite{Mackinlay}. Movement–concept mappings were chosen to enhance semantic congruence between physical action and data and numeracy concept. For example, extending the right elbow to the right mirrors the visual rightward tail of a right-skewed distribution. Raising a knee 'n' times encodes the number 'n' as a direct embodied count. Mapping candidates were identified through a formative ideation session and refined through informal think-aloud testing before the main study. 
\begin{table}[ht]
  \centering
  \caption{Samples of Kinetiq’s data literacy movement games}
  \label{tab:minigames}
  \begin{tabular}{|p{4cm}|p{4cm}|p{5cm}|p{5cm}|}
    \hline
    \textbf{Mini Game} & \textbf{Required Movements} \\
    \hline
    Catch or Head the Central Tendency & Reach or slap to catch or head floating bubbles with correct means, medians, or modes, while dodging incorrect ones. \\
    \hline
    Elbow distribution skews & Extend arms and  elbows to hit target answers that correspond physically and conceptually.
\\
    \hline
    Knee mixed problems & Raise legs and knees. Repeat movements to represent the correct number.
 \\
    \hline
    Ditch outliers & Identify and move outliers out of the box \\
   \hline
Coordinate Quest & Navigate the coordinate grid with arm movement.\\
    \hline
    Collaborative (Multiplayer) & Mixed movement and data problem-solving \\
    \hline
    
  \end{tabular}
\end{table}
\begin{figure}[h]
  \centering
   \begin{minipage}[b]{0.45\textwidth}
    {\includegraphics[width=\textwidth]{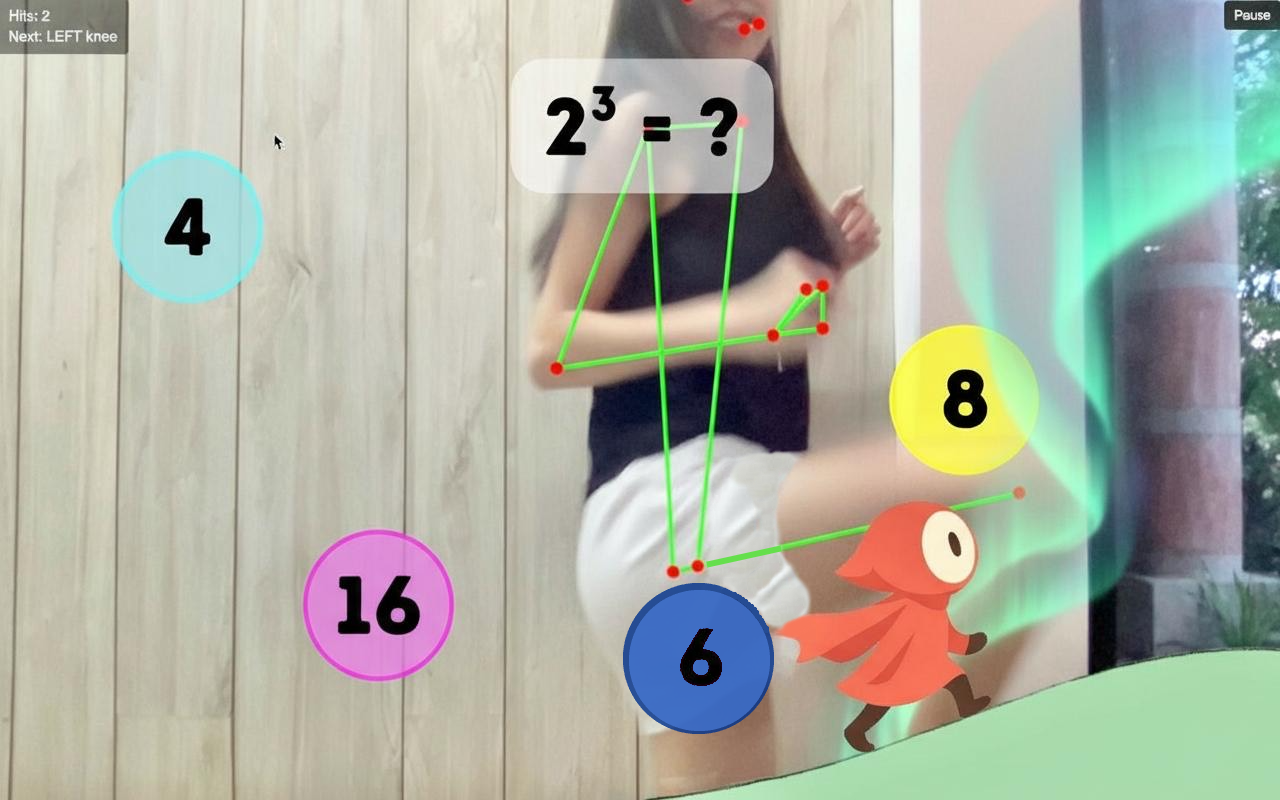}} 
    \caption{Kinetiq affords multiple gestures such as knee lifts to answer data and numeracy questions; the character walks uphill as the learner progresses. }
    \end{minipage}
  \hfill
  \begin{minipage}[b]{0.45\textwidth}
    {\includegraphics[width=\textwidth]{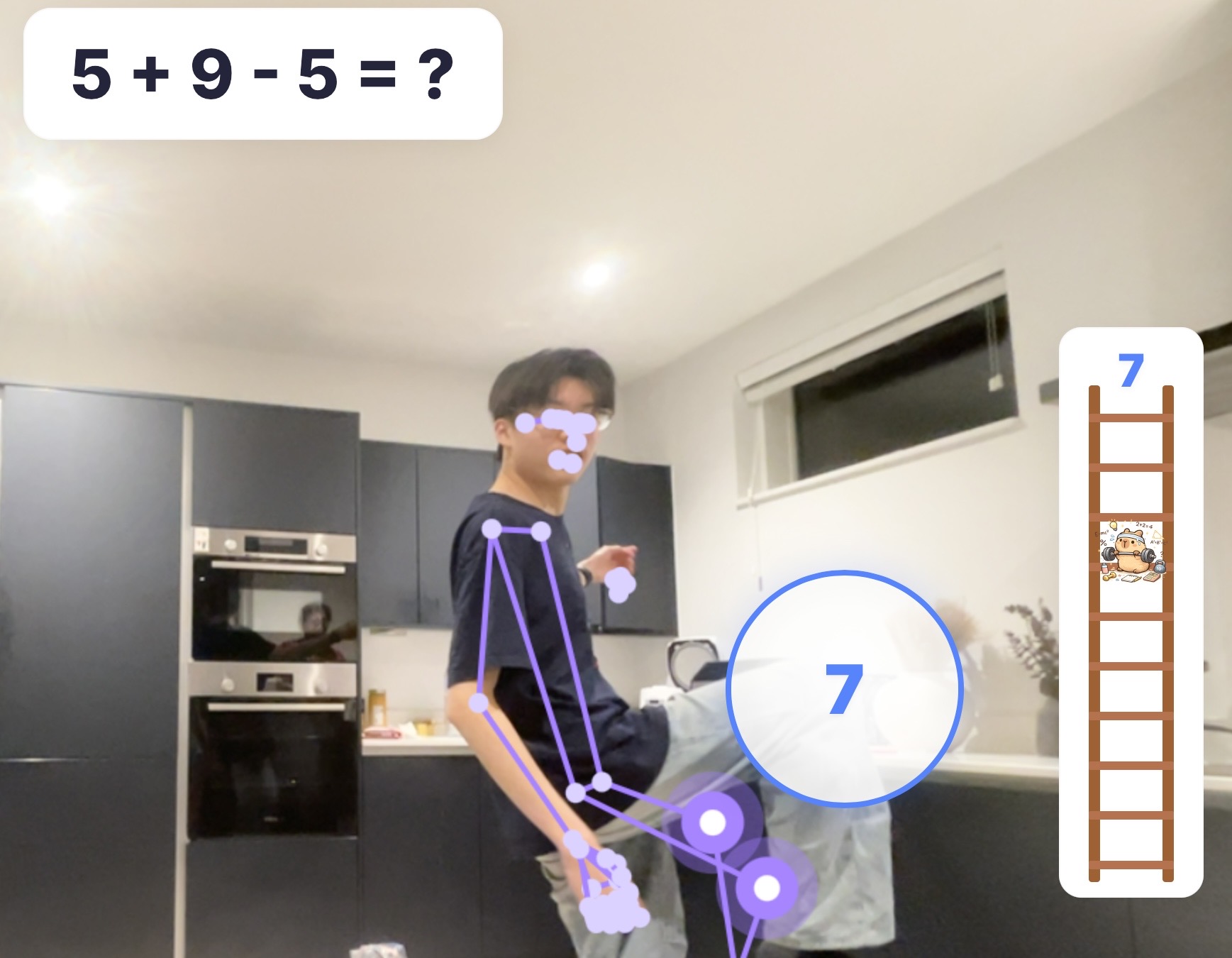}} 
  \caption{Users answer by repeated knee lifts, with the number of lifts matching their intended answer. The character climbs up as the learner gains mastery.}
  \end{minipage}
  \Description{}
\end{figure}
\section{Preliminary User Study}
Sixteen participants (9 female, 7 male) participated in a preliminary within-subjects study to examine how embodied micro-movements shape engagement with foundational data literacy tasks. Participants engaged in three 15-minute activity sessions of using Kinetiq, IXL, and Khan Academy. All tasks covered the same data concepts to ensure equivalence in content. The order of platforms was counterbalanced across participants. Participants were informed that all video processing occurred on-device and that no webcam data was recorded or transmitted beyond the session. Seated and low-amplitude gesture alternatives were available to any participant who required them. Sessions were separated by a 5-minute rest break to reduce carryover fatigue.
We measured affective valence: The Feeling Scale (FS) was used to assess affective state. We also measured enjoyment and collected qualitative feedback on usability, happiness, motivation, and willingness to reuse the platforms. Researchers took note of facial expressions, physical engagement, and posture during use.

\textbf{Findings}
Affective valence scores were significantly higher during Kinetiq sessions compared to both IXL and Khan Academy (mean valence +2.4 points, p < .05), with no significant difference between the two conventional platforms. Enjoyment scores were higher after Kinetiq sessions and lower after the other platforms, with participants frequently describing it as “very fun”. Ninety four percent of participants reported feeling more motivated to return to Kinetiq than to either comparison tool. They also reported higher happiness after using Kinetiq (M = 6.0, SD = 0.76) than before (M=5.0, SD = 1.07). A one-tailed paired-samples t-test confirmed that this increase was statistically significant (p<0.01). Results also show that they preferred Kinetiq (M=7.3) significantly more than Khan Academy (M=5.2, p<0.01). All participants show high levels of enthusiasm and excitement and reported feeling “refreshed” and “awake” after using Kinetiq, similar to the feeling after light physical exercise. Behaviorally, participants sat more upright, smiled, and showed engaged postures, in contrast to the slouched posture observed during traditional platform usage.
We also administered a brief pre- and post-assessment of 10 data literacy items across all three conditions. Participants using Kinetiq demonstrated learning gains comparable to those using IXL and Khan Academy, with a trend favoring Kinetiq. These findings suggest that embodied interaction does not impede learning and may support the learning of data and numeracy concepts through physical grounding while promoting engagement, motivation, and fun. We plan to test with larger samples and longer interventions.
\subsection{Implications for Data Literacy}
Our results, while preliminary, suggest several implications for the data literacy community. Although data literacy concepts are often perceived as dry or abstract, the significant improvements in affect, enjoyment, and motivation we observed suggest that movement-integrated interaction could help sustain engagement with data literacy practice.
Second, physical grounding may support data reasoning. While our current study focused more on affective and motivational outcomes, embodied cognition theory and prior empirical work suggest that physically interacting with data concepts such as gesturing with the left elbow to represent a left-skewed distribution may help learners build richer mental representations than selecting answers by clicking alone. Future work should directly measure whether movement-integrated data literacy tasks improve conceptual understanding and memory.

Our system and study also open up broader questions for the data literacy community. One key question is whether physical interaction changes how people reason about statistical concepts, beyond simply making learning more engaging. A related question is whether movement-based interaction could serve as an alternative assessment modality for data literacy; for example, allowing researchers and educators to infer understanding from how learners move, reach, or spatially organize data rather than only from their multiple-choice responses. Finally, this approach raises questions about how embodied, playful interaction might address anxiety, which is a well-documented barrier to learning in this domain. Together, these questions suggest the need for future work that not only measures affective and learning outcomes, but also examines how movement shapes reasoning, assessment, and emotional experience in data literacy learning. 
\begin{figure}[h]
\includegraphics[width=0.5\linewidth]{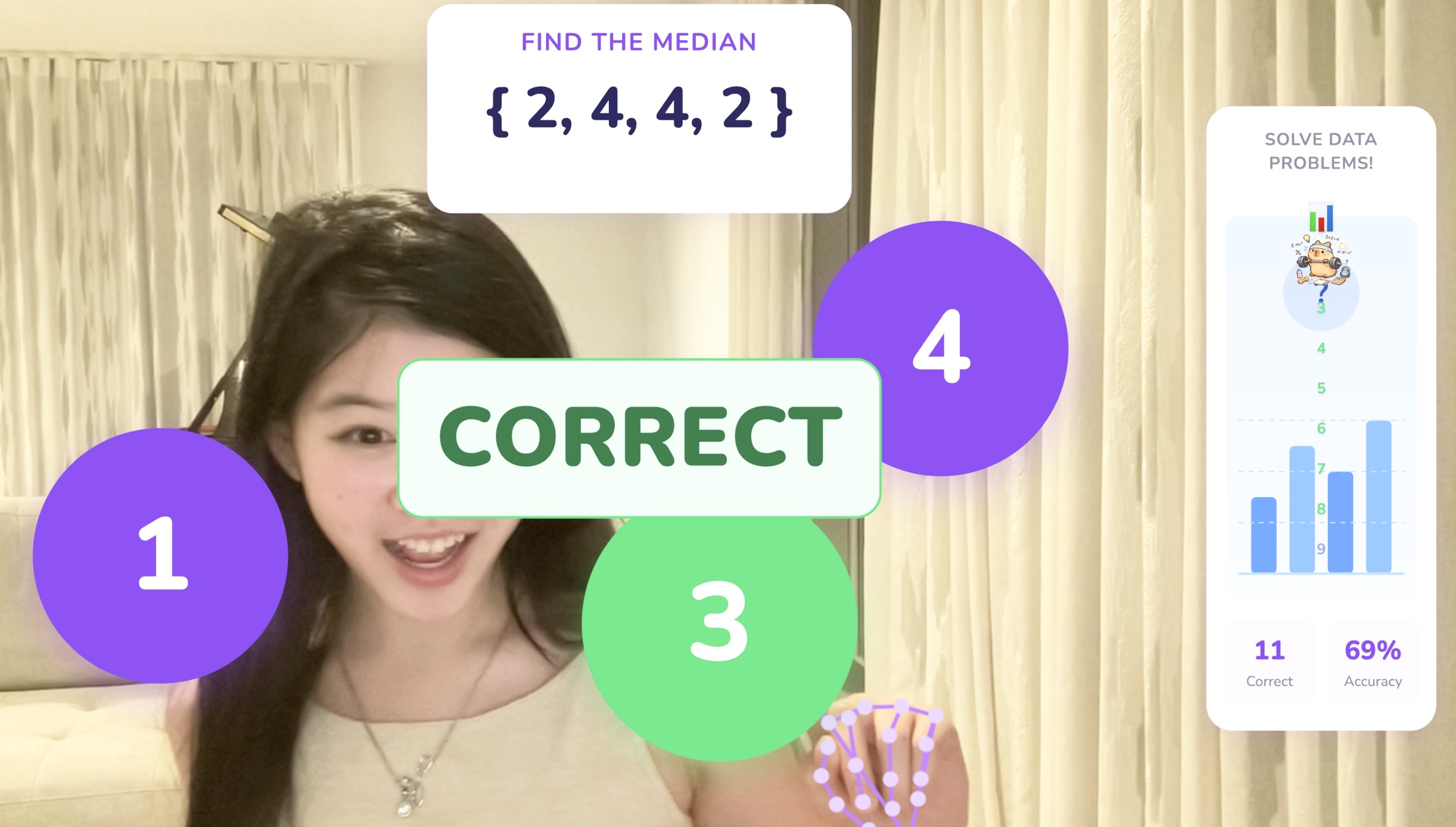}
  \caption{Data literacy practice}
  \Description{Learning and Getting Healthier, one move at a time.}
  \label{fig:teaser}
\end{figure}

Our exploration of Kinetiq also introduces a rich set of unresolved design questions for embodied data literacy systems that extend beyond this single prototype. These include representational choices such as whether learners should see themselves via live video or interact through personalized avatars (in our study, all participants preferred seeing themselves, although avatars may offer benefits for privacy, accessibility, and creative abstraction). We also encountered trade-offs in visual designs, including whether data elements (e.g. bubbles) should share a uniform color or vary in color. Spatial and temporal design decisions likewise remain open: how large should targets and text be to balance accuracy, readability, and challenge; and should targets move slowly to support reflection or more rapidly to enhance kinetic flow and engagement? In addition, the optimal temporal structure of embodied data literacy training sessions remains an open question. Should a session conclude when learners achieve a predefined number of correct target points, hence emphasizing mastery, or when a fixed time limit is reached, prioritizing pacing and sustained engagement and fun? Understanding how these alternatives shape attention, fatigue, and  learning outcomes is an important direction for future design and empirical investigation. The affective environment such as the role of music and sound also raises questions about how audio might reduce data and statistics anxiety in embodied learning contexts or distract from reasoning. Together, these questions suggest a broader design space in which embodiment, accessibility, expressiveness, and effectiveness must be carefully considered for future movement-based data literacy learning environments.
\section{Conclusion and Future Work}

We introduce Kinetiq, a web-based and cross-platform learning system that integrates body movement with data problem solving. Kinetiq reimagines digital learning by showing how small, intentional movements can shift data literacy practice from passive interaction toward more joyful, attentive, and resilient engagement. Our contributions lie in (1) a new class of AI-mediated, naturalistic movement microinterventions for learning and wellbeing, (2) a low-barrier interaction system that enables full-body activity in small spaces without wearables, sensors, or calibration, (3) a new way of thinking about data literacy that links learning, movement, and wellbeing, and (4) empirical evidence that such natural gestures in movement-based learning improve affect, enjoyment, and motivation during digital data problem-solving. Kinetiq thus invites a future in which learning is not separate from movement or the environment, but instead is thoughtfully and pleasurably connected to both. Rather than treating movement as a break from learning, Kinetiq treats movement as the interface for learning. We demonstrate that embodied interaction offers a promising new modality for developing data literacy competencies. This preliminary study involved a limited sample, primarily focused on short-term outcomes. Further longitudinal studies are needed to measure sustained effects. Future work will also explore expanding data literacy content to include data visualization interpretation, probability reasoning, and critical evaluation of data-driven claims. Additional directions include user-specific adaptations for neurodiverse learners and older adults.
We invite discussion on how physical interaction modalities might reshape the teaching and learning of data literacy, and how assessment frameworks might be extended to capture learning that occurs through the body in addition to through the mind.

\begin{acks}
We thank Stanford Medical School instructors and affiliates for feedback on the earliest technical and design prototype variant.
\end{acks}

\section{Selection and Participation of Children}
Sixteen participants (9 female, 7 male, age range 5-16) took part in preliminary feedback sessions recruited through local community networks. We acknowledge that the sample may not reflect broader demographic diversity, and future work will aim to recruit more diverse participants across socioeconomic and cultural backgrounds. Parents provided informed consent and confirmed their understanding that sessions involved movement-based interaction with a learning system, that all video processing occurred on-device, and that no video data would be retained or shared. Participants benefited from exposure to a novel, engaging approach to data literacy learning. Seated and low-amplitude gesture alternatives were available to any participant who required them. Contact information was collected solely for the purpose of sharing findings with participants and their families upon publication and will not be used for any other purpose. Participants and their families will be informed of findings upon publication.

\bibliographystyle{ACM-Reference-Format}
\bibliography{sample-base}


\begin{thebibliography}{21}


\ifx \showCODEN    \undefined \def \showCODEN     #1{\unskip}     \fi
\ifx \showISBNx    \undefined \def \showISBNx     #1{\unskip}     \fi
\ifx \showISBNxiii \undefined \def \showISBNxiii  #1{\unskip}     \fi
\ifx \showISSN     \undefined \def \showISSN      #1{\unskip}     \fi
\ifx \showLCCN     \undefined \def \showLCCN      #1{\unskip}     \fi
\ifx \shownote     \undefined \def \shownote      #1{#1}          \fi
\ifx \showarticletitle \undefined \def \showarticletitle #1{#1}   \fi
\ifx \showURL      \undefined \def \showURL       {\relax}        \fi
\providecommand\bibfield[2]{#2}
\providecommand\bibinfo[2]{#2}
\providecommand\natexlab[1]{#1}
\providecommand\showeprint[2][]{arXiv:#2}

\bibitem[Burckhardt et~al\mbox{.}(2021)]%
        {Burckhardt01012021}
\bibfield{author}{\bibinfo{person}{Philipp Burckhardt}, \bibinfo{person}{Rebecca Nugent}, {and} \bibinfo{person}{Christopher~R. Genovese}.} \bibinfo{year}{2021}\natexlab{}.
\newblock \showarticletitle{Teaching Statistical Concepts and Modern Data Analysis With a Computing-Integrated Learning Environment}.
\newblock \bibinfo{journal}{\emph{Journal of Statistics and Data Science Education}} \bibinfo{volume}{29}, \bibinfo{number}{sup1} (\bibinfo{year}{2021}), \bibinfo{pages}{S61--S73}.
\newblock
\showeprint{https://doi.org/10.1080/10691898.2020.1854637}
\href{https://doi.org/10.1080/10691898.2020.1854637}{doi:\nolinkurl{10.1080/10691898.2020.1854637}}


\bibitem[Chu et~al\mbox{.}(2025)]%
        {Chu2025}
\bibfield{author}{\bibinfo{person}{Junyi Chu}, \bibinfo{person}{Kristine Zheng}, {and} \bibinfo{person}{Judith~E Fan}.} \bibinfo{year}{2025}\natexlab{}.
\newblock \showarticletitle{What makes people think a puzzle is fun to solve?}
\newblock \bibinfo{journal}{\emph{Proceedings of the Annual Meeting of the Cognitive Science Society}} \bibinfo{volume}{47}, \bibinfo{number}{0} (\bibinfo{year}{2025}).
\newblock


\bibitem[Corliss(2024)]%
        {harvardMuchSit}
\bibfield{author}{\bibinfo{person}{Julie Corliss}.} \bibinfo{year}{2024}\natexlab{}.
\newblock \bibinfo{title}{{H}ow much do you sit, stand, and move each day? - {H}arvard {H}ealth --- health.harvard.edu}.
\newblock \bibinfo{howpublished}{\url{https://www.health.harvard.edu/heart-health/how-much-do-you-sit-stand-and-move-each-day}}.
\newblock


\bibitem[Duran et~al\mbox{.}(2023)]%
        {duran2023breaking}
\bibfield{author}{\bibinfo{person}{Andrea~T. Duran}, \bibinfo{person}{Ciaran~P. Friel}, \bibinfo{person}{Maria~A. Serafini}, \bibinfo{person}{Ipek Ensari}, \bibinfo{person}{Ying~Kuen Cheung}, {and} \bibinfo{person}{Keith~M. Diaz}.} \bibinfo{year}{2023}\natexlab{}.
\newblock \showarticletitle{Breaking Up Prolonged Sitting to Improve Cardiometabolic Risk: Dose–Response Analysis of a Randomized Crossover Trial}.
\newblock \bibinfo{journal}{\emph{Medicine \& Science in Sports \& Exercise}} \bibinfo{volume}{55}, \bibinfo{number}{5} (\bibinfo{date}{May} \bibinfo{year}{2023}), \bibinfo{pages}{847--855}.
\newblock
\href{https://doi.org/10.1249/MSS.0000000000003109}{doi:\nolinkurl{10.1249/MSS.0000000000003109}}


\bibitem[Fan et~al\mbox{.}(2024)]%
        {Fan2024}
\bibfield{author}{\bibinfo{person}{Danyang Fan}, \bibinfo{person}{Gene S-H Kim}, \bibinfo{person}{Olivia Tomassetti}, \bibinfo{person}{Shloke~Nirav Patel}, \bibinfo{person}{Sile O'Modhrain}, \bibinfo{person}{Victor~R Lee}, {and} \bibinfo{person}{Sean Follmer}.} \bibinfo{year}{2024}\natexlab{}.
\newblock \showarticletitle{Tangible Stats: An Embodied and Multimodal Platform for Teaching Data and Statistics to Blind and Low Vision Students}. In \bibinfo{booktitle}{\emph{Extended Abstracts of the CHI Conference on Human Factors in Computing Systems}} (Honolulu, HI, USA) \emph{(\bibinfo{series}{CHI EA '24})}. \bibinfo{publisher}{Association for Computing Machinery}, \bibinfo{address}{New York, NY, USA}, Article \bibinfo{articleno}{310}, \bibinfo{numpages}{9}~pages.
\newblock
\showISBNx{9798400703317}
\href{https://doi.org/10.1145/3613905.3650793}{doi:\nolinkurl{10.1145/3613905.3650793}}


\bibitem[Habgood and Ainsworth(2011)]%
        {habgood2011motivation}
\bibfield{author}{\bibinfo{person}{Matthew P.~J. Habgood} {and} \bibinfo{person}{Shaaron~E. Ainsworth}.} \bibinfo{year}{2011}\natexlab{}.
\newblock \showarticletitle{Motivating children to learn effectively: Exploring the value of intrinsic integration in educational games}.
\newblock \bibinfo{journal}{\emph{Journal of the Learning Sciences}} \bibinfo{volume}{20}, \bibinfo{number}{2} (\bibinfo{year}{2011}), \bibinfo{pages}{169--206}.
\newblock
\href{https://doi.org/10.1080/10508406.2010.508029}{doi:\nolinkurl{10.1080/10508406.2010.508029}}


\bibitem[Hunicke et~al\mbox{.}(2004)]%
        {hunicke2004mda}
\bibfield{author}{\bibinfo{person}{Robin Hunicke}, \bibinfo{person}{Marc LeBlanc}, {and} \bibinfo{person}{Robert Zubek}.} \bibinfo{year}{2004}\natexlab{}.
\newblock \showarticletitle{MDA: A Formal Approach to Game Design and Game Research}. In \bibinfo{booktitle}{\emph{Proceedings of the AAAI Workshop on Challenges in Game AI}}, Vol.~\bibinfo{volume}{4}. \bibinfo{pages}{1--5}.
\newblock


\bibitem[Jason(2017)]%
        {harvard2017}
\bibfield{author}{\bibinfo{person}{Zachary Jason}.} \bibinfo{year}{2017}\natexlab{}.
\newblock \bibinfo{title}{{B}ored {O}ut of {T}heir {M}inds}.
\newblock \bibinfo{howpublished}{\url{https://www.gse.harvard.edu/ideas/ed-magazine/17/01/bored-out-their-minds}}.
\newblock
\newblock
\shownote{[Accessed 12-07-2025]}.


\bibitem[Li and Kangas(2024)]%
        {Li2024}
\bibfield{author}{\bibinfo{person}{Xiaoyan Li} {and} \bibinfo{person}{Marjaana Kangas}.} \bibinfo{year}{2024}\natexlab{}.
\newblock \showarticletitle{A systematic literature review of playful learning in primary education: teachers’ pedagogical activities}.
\newblock \bibinfo{journal}{\emph{Education 3-13}} \bibinfo{volume}{0}, \bibinfo{number}{0} (\bibinfo{year}{2024}), \bibinfo{pages}{1--16}.
\newblock
\showeprint{https://doi.org/10.1080/03004279.2024.2416954}
\href{https://doi.org/10.1080/03004279.2024.2416954}{doi:\nolinkurl{10.1080/03004279.2024.2416954}}


\bibitem[Mackinlay(1986)]%
        {Mackinlay}
\bibfield{author}{\bibinfo{person}{Jock Mackinlay}.} \bibinfo{year}{1986}\natexlab{}.
\newblock \showarticletitle{Automating the design of graphical presentations of relational information}.
\newblock \bibinfo{journal}{\emph{ACM Trans. Graph.}} \bibinfo{volume}{5}, \bibinfo{number}{2} (\bibinfo{date}{April} \bibinfo{year}{1986}), \bibinfo{pages}{110–141}.
\newblock
\showISSN{0730-0301}
\href{https://doi.org/10.1145/22949.22950}{doi:\nolinkurl{10.1145/22949.22950}}


\bibitem[McGonigal(2019)]%
        {mcgonigal2019}
\bibfield{author}{\bibinfo{person}{Kelly McGonigal}.} \bibinfo{year}{2019}\natexlab{}.
\newblock \bibinfo{booktitle}{\emph{The joy of movement: How exercise helps us find happiness, hope, connection, and courage}}.
\newblock \bibinfo{publisher}{Penguin}.
\newblock


\bibitem[Meng(2009)]%
        {Meng}
\bibfield{author}{\bibinfo{person}{Xiao-Li Meng}.} \bibinfo{year}{2009}\natexlab{}.
\newblock \bibinfo{title}{{S}tatistics: {Y}our chance for happiness (or misery) | {D}epartment of {S}tatistics --- statistics.fas.harvard.edu}.
\newblock \bibinfo{howpublished}{\url{https://statistics.fas.harvard.edu/statistics-your-chance-happiness-or-misery}}.
\newblock
\newblock
\shownote{[Accessed 07-02-2026]}.


\bibitem[Novack and Goldin-Meadow(2015)]%
        {novack2015}
\bibfield{author}{\bibinfo{person}{Miriam Novack} {and} \bibinfo{person}{Susan Goldin-Meadow}.} \bibinfo{year}{2015}\natexlab{}.
\newblock \showarticletitle{Learning from gesture: How our hands change our minds}.
\newblock \bibinfo{journal}{\emph{Educational Psychology Review}} \bibinfo{volume}{27}, \bibinfo{number}{3} (\bibinfo{date}{Sept.} \bibinfo{year}{2015}), \bibinfo{pages}{405--412}.
\newblock
\href{https://doi.org/10.1007/s10648-015-9325-3}{doi:\nolinkurl{10.1007/s10648-015-9325-3}}


\bibitem[Onwuegbuzie and Leech(2003)]%
        {Onwuegbuzie2003AssessmentIS}
\bibfield{author}{\bibinfo{person}{Anthony~John Onwuegbuzie} {and} \bibinfo{person}{Nancy~L. Leech}.} \bibinfo{year}{2003}\natexlab{}.
\newblock \showarticletitle{Assessment in Statistics Courses: More than a tool for evaluation}.
\newblock \bibinfo{journal}{\emph{Assessment \& Evaluation in Higher Education}}  \bibinfo{volume}{28} (\bibinfo{year}{2003}), \bibinfo{pages}{115 -- 127}.
\newblock
\urldef\tempurl%
\url{https://api.semanticscholar.org/CorpusID:143587956}
\showURL{%
\tempurl}


\bibitem[Organization(2020)]%
        {who2020}
\bibfield{author}{\bibinfo{person}{World~Health Organization}.} \bibinfo{year}{2020}\natexlab{}.
\newblock \bibinfo{title}{Who Guidelines on Physical Activity and Sedentary Behavior}.
\newblock \bibinfo{howpublished}{\url{https://iris.who.int/bitstream/handle/10665/336656/9789240015128-eng.pdf?sequence=1}}.
\newblock
\newblock
\shownote{[Accessed 12-07-2025]}.


\bibitem[Schmidt et~al\mbox{.}(2016)]%
        {schmidt2016executive}
\bibfield{author}{\bibinfo{person}{Mirko Schmidt}, \bibinfo{person}{Valentin Benzing}, {and} \bibinfo{person}{Mario Kamer}.} \bibinfo{year}{2016}\natexlab{}.
\newblock \showarticletitle{Classroom-based physical activity improves children’s executive function}.
\newblock \bibinfo{journal}{\emph{PLoS ONE}} \bibinfo{volume}{11}, \bibinfo{number}{8} (\bibinfo{year}{2016}), \bibinfo{pages}{e0167501}.
\newblock
\href{https://doi.org/10.1371/journal.pone.0167501}{doi:\nolinkurl{10.1371/journal.pone.0167501}}


\bibitem[Shapiro and Spaulding(2025)]%
        {shapiro2025}
\bibfield{author}{\bibinfo{person}{Lawrence Shapiro} {and} \bibinfo{person}{Shannon Spaulding}.} \bibinfo{year}{2025}\natexlab{}.
\newblock \showarticletitle{{Embodied Cognition}}.
\newblock In \bibinfo{booktitle}{\emph{The {Stanford} Encyclopedia of Philosophy} (\bibinfo{edition}{{S}ummer 2025} ed.)}, \bibfield{editor}{\bibinfo{person}{Edward~N. Zalta} {and} \bibinfo{person}{Uri Nodelman}} (Eds.). \bibinfo{publisher}{Metaphysics Research Lab, Stanford University}.
\newblock


\bibitem[Smith et~al\mbox{.}(2024)]%
        {smith2024designing}
\bibfield{author}{\bibinfo{person}{C.~E. Smith}, \bibinfo{person}{S. Lee}, \bibinfo{person}{T.~D. Allen}, \bibinfo{person}{M.~L. Wallace}, \bibinfo{person}{R. Andel}, \bibinfo{person}{O.~M. Buxton}, \bibinfo{person}{S.~R. Patel}, {and} \bibinfo{person}{D.~M. Almeida}.} \bibinfo{year}{2024}\natexlab{}.
\newblock \showarticletitle{Designing work for healthy sleep: A multidimensional, latent transition approach to employee sleep health}.
\newblock \bibinfo{journal}{\emph{Journal of Occupational Health Psychology}} \bibinfo{volume}{29}, \bibinfo{number}{6} (\bibinfo{year}{2024}), \bibinfo{pages}{409--430}.
\newblock
\href{https://doi.org/10.1037/ocp0000375}{doi:\nolinkurl{10.1037/ocp0000375}}


\bibitem[Trustpilot({[n.\,d.]})]%
        {Trustpilot}
\bibfield{author}{\bibinfo{person}{Trustpilot}.} \bibinfo{year}{[n.\,d.]}\natexlab{}.
\newblock \bibinfo{title}{Trustpilot Reviews: {E}xperience the power of customer reviews --- trustpilot.com}.
\newblock
\urldef\tempurl%
\url{https://www.trustpilot.com}
\showURL{%
\tempurl}
\newblock
\shownote{[Accessed 07-02-2026]}.


\bibitem[Vazou et~al\mbox{.}(2020)]%
        {vazou2020cognition}
\bibfield{author}{\bibinfo{person}{Spiridoula Vazou}, \bibinfo{person}{Caterina Pesce}, \bibinfo{person}{Kimberley~D Lakes}, {and} \bibinfo{person}{Ann Smiley-Oyen}.} \bibinfo{year}{2020}\natexlab{}.
\newblock \showarticletitle{More than one road leads to Rome: A narrative review and meta-analysis of physical activity intervention effects on children’s cognition}.
\newblock \bibinfo{journal}{\emph{Frontiers in Psychology}}  \bibinfo{volume}{9} (\bibinfo{year}{2020}), \bibinfo{pages}{2079}.
\newblock
\href{https://doi.org/10.3389/fpsyg.2018.02079}{doi:\nolinkurl{10.3389/fpsyg.2018.02079}}


\bibitem[Wilson(2002)]%
        {wilson2002}
\bibfield{author}{\bibinfo{person}{Margaret Wilson}.} \bibinfo{year}{2002}\natexlab{}.
\newblock \showarticletitle{Six views of embodied cognition}.
\newblock \bibinfo{journal}{\emph{Psychonomic Bulletin \& Review}} \bibinfo{volume}{9}, \bibinfo{number}{4} (\bibinfo{year}{2002}), \bibinfo{pages}{625--636}.
\newblock
\href{https://doi.org/10.3758/BF03196322}{doi:\nolinkurl{10.3758/BF03196322}}


\end{thebibliography}






\end{document}